\documentclass[preprint,12pt]{elsarticle}

\usepackage{graphicx}

\usepackage{amssymb}
\usepackage{amsmath}
\usepackage{multirow}
\usepackage[font=small,labelfont=bf]{caption}
\DeclareCaptionFont{tiny}{\scriptsize}
\DeclareCaptionFont{tinytiny}{\tiny}

\biboptions{sort&compress}

\newlength{\dslashwidth}

\newcommand{\beq}{\begin{equation}} 
\newcommand{\eeq}{\end{equation}}
\newcommand{\beqa}{\begin{eqnarray}} 
\newcommand{\eeqa}{\end{eqnarray}}
\newcommand{\newc}{\newcommand}
\newcommand{\bq}{\begin{equation}}
\newcommand{\eq}{\end{equation}}
\newcommand{\ba}{\begin{array}}
\newcommand{\ea}{\end{array}}
\newcommand{\bqa}{\begin{eqnarray}}
\newcommand{\eqa}{\end{eqnarray}}

\newcommand{\lnf}{{\ifmmode \Lambda^{(N_f)} \else $\Lambda^{(N_f)}$\fi}}
\newcommand{\ms}{{\ifmmode \overline{MS} \else $\overline{MS}$\fi}}
\newcommand{\dr}{{\ifmmode \overline{DR} \else $\overline{DR}$\fi}}
\newcommand{\lms}{{\ifmmode \Lambda^{(5)}_{\overline{MS}} \else $\Lambda^{(5)}_{\overline{MS}}$\fi}}
\newcommand{\lam}{{\ifmmode \Lambda \else $\Lambda$\fi}}
\newcommand{\mev}{{\ifmmode {\rm MeV} \else ${\rm MeV}$\fi}}
\newcommand{\gev}{{\ifmmode {\rm GeV} \else ${\rm GeV}$\fi}}
\newcommand{\gevc}{{\ifmmode {\rm GeV/c^2} \else ${\rm GeV/c^2}$\fi}}
\newcommand{\tev}{{\ifmmode {\rm TeV} \else ${\rm TeV}$\fi}}
\newcommand{\tevc}{{\ifmmode {\rm TeV/c^2} \else ${\rm TeV/c^2}$\fi}}
\newcommand{\lp}{{\ifmmode L^+  \else $L^+$\fi}}
\newcommand{\lm}{{\ifmmode L^-  \else $L^-$\fi}}
\newcommand{\mlp}{{\ifmmode M(L^-) \else $M(L^-)$\fi}}
\newcommand{\mlz}{{\ifmmode M(L^0) \else $M(L^0)$\fi}}
\newcommand{\lz}{{\ifmmode L^0 \else $L^0$\fi}}
\newcommand{\ev}{{\ifmmode GeV/c^2 \else $GeV/c^2$\fi}}
\newcommand{\tri}{{\ifmmode \triangleup \else $\triangleup$\fi}}
\newcommand{\unl}{{\ifmmode U_{lL^0} \else $U_{lL^0}$\fi}}\newcommand{\gL}{{\ifmmode g_L \else $g_{L}$\fi}}
\newcommand{\gR}{{\ifmmode g_R  \else $g_{R}$\fi}}
\newcommand{\gumu}{{\ifmmode \gamma^{\mu} \else $\gamma^{\mu}$\fi}}
\newcommand{\gunu}{{\ifmmode \gamma^{\nu} \else $\gamma^{\nu}$\fi}}
\newcommand{\gdmu}{{\ifmmode \gamma_{\mu} \else $\gamma_{\mu}$\fi}}
\newcommand{\gdnu}{{\ifmmode \gamma_{\nu} \else $\gamma_{\nu}$\fi}}
\newcommand{\stw}{{\ifmmode\sin^2\theta_W \else $\sin^{2}\theta_{W}$ \fi}}
\newcommand{\sws}{{\ifmmode \;\sin^2\theta_W  \else $\;\sin^{2}\theta_{W}$ \fi}}
\newcommand{\cws}{{\ifmmode \;\cos^2\theta_W  \else $\;\cos^{2}\theta_{W}$ \fi}}
\newcommand{\sw}{{\ifmmode \;\sin\theta_W  \else $\sin\theta_{W}$ \fi}}
\newcommand{\cw}{{\ifmmode \;\cos\theta_W  \else $\;\cos\theta_{W}$ \fi}}
\newcommand{\tw}{{\ifmmode \;\tan\theta_W  \else $\;\tan\theta_{W}$ \fi}}
\newcommand{\qq}{{\ifmmode q\overline{q} \else $q\overline{q}$\fi}}
\newcommand{\lR}{{\ifmmode l_R \else $l_R$\fi}}
\newcommand{\lL}{{\ifmmode l_L \else $l_L$\fi}}
\newcommand{\nt}{{\ifmmode \nu_{\tau} \else $\nu_{\tau}$\fi}}
\newcommand{\nuR}{{\ifmmode \nu_R  \else $\nu_R$\fi}}
\newcommand{\nuL}{{\ifmmode \nu_L  \else $\nu_L$\fi}}
\newcommand{\qR}{{\ifmmode g_R  \else $q_R$\fi}}
\newcommand{\qL}{{\ifmmode q_L  \else $q_L$\fi}}
\newcommand{\qRp}{{\ifmmode q_R'  \else $q_{R}$'\fi}}
\newcommand{\qLp}{{\ifmmode q_L'  \else $q_{L}$'\fi}}
\newcommand{\est}{{\ifmmode e^{\bf \ast} \else $e^{\bf \ast}$\fi}}
\newcommand{\lst}{{\ifmmode l^{\bf \ast} \else $l^{\bf \ast}$\fi}}
\newcommand{\must}{{\ifmmode \mu^{\bf \ast} \else $\mu^{\bf \ast}$\fi}}
\newcommand{\taust}{{\ifmmode \tau^{\bf \ast} \else $\tau^{\bf \ast}$ \fi}}
\newcommand{\pperp}{{\ifmmode p_t  \else $p_t$\fi}}
\newcommand{\et}{{\ifmmode E_t  \else $E_t$\fi}}
\newcommand{\xt}{{\ifmmode x_t  \else $x_t$\fi}}
\newcommand{\smumu}{{\ifmmode \sigma_{\mu\mu}  \else $\sigma_{\mu\mu}$ \fi}}
\newcommand{\eg}{{\ifmmode e\gamma  \else $e\gamma$\fi}}
\newcommand{\epem}{{\ifmmode e^+e^-  \else $e^+e^-$\fi}}
\newcommand{\lplm}{{\ifmmode L^+L^-  \else $L^+L^-$\fi}}
\newcommand{\pp}{{\ifmmode p\overline p  \else $p\overline p$\fi}}
\newcommand{\llz}{{\ifmmode L^0\overline{L}^0 \else $L^0\overline{L}^0$\fi}}
\newcommand{\epemt}{{\ifmmode e^+e^- \to  \else $e^+e^- \to$\fi}}
\newcommand{\eb}{{\ifmmode E_{beam}  \else $E_{beam}$\fi}}
\newcommand{\ip}{{\ifmmode pb^{-1}  \else $pb^{-1}$\fi}}
\newcommand{\upm}{{\ifmmode ^{\pm}  \else $^{\pm}$\fi}}
\newcommand{\de}{{\ifmmode ^{\circ}  \else $^{\circ}$ \fi}}
\newcommand{\appr}{{\ifmmode \sim \else $\sim$ \fi}}
\newcommand{\corresp}{{\ifmmode \stackrel{\wedge}{=} \else $\stackrel{\wedge}{=}$ \fi}}
\newcommand{\sqrts}{{\ifmmode \sqrt{s} \else $\sqrt{s}$\fi}}
\newcommand{\zz}{{\ifmmode Z^0  \else $Z^0$\fi}}
\newcommand{\mz}{{\ifmmode M_{Z}  \else $M_{Z}$\fi}}
\newcommand{\mzs}{{\ifmmode M_{Z}^2  \else $M_{Z}^2$\fi}}
\newcommand{\mw}{{\ifmmode M_{W}  \else $M_{W}$\fi}}
\newcommand{\mws}{{\ifmmode M_{W}^2  \else $M_{W}^2$\fi}}
\newcommand{\mh}{{\ifmmode M_{Higgs}  \else $M_{Higgs}$\fi}}
\newcommand{\msusy}{{\ifmmode M_{SUSY}  \else $M_{SUSY}$\fi}}
\newcommand{\msusys}{{\ifmmode M_{SUSY}^2  \else $M_{SUSY}^2$\fi}}
\newcommand{\su}{{\ifmmode SU(3)_C\otimes\- SU(2)_L\otimes\- U(1)_Y \else $SU(3)_C\otimes\A0SU(2)_L\otimes U(1)_Y$\fi}}
\newcommand{\suthree}{{\ifmmode SU(3)_C  \else $SU(3)_C$\fi}}
\newcommand{\sutwo}{{\ifmmode  SU(2)_L\otimes U(1)_Y \else $SU(2)_L\otimes U(1)_Y$\fi}}
\newcommand{\taup}{{\ifmmode \tau_{proton} \else $\tau_{proton}$\fi}}
\newcommand{\as}{{\ifmmode \alpha_{s}  \else $\alpha_{s}$\fi}}
\newcommand{\mgut}{{\ifmmode M_{GUT}  \else $M_{GUT}$\fi}}
\newcommand{\mguts}{{\ifmmode M_{GUT}^2  \else $M_{GUT}^2$\fi}}
\newcommand{\mzero}{{\ifmmode m_0        \else $m_0$\fi}}
\newcommand{\mhalf}{{\ifmmode m_{1/2}    \else $m_{1/2}$\fi}}
\newcommand{\sq}{{\ifmmode \tilde{q}    \else $\tilde{q}$\fi}}
\newcommand{\gl}{{\ifmmode \tilde{g}    \else $\tilde{g}$\fi}}
\newcommand{\mb}{{\ifmmode m_{b}    \else $m_{b}$\fi}}
\newcommand{\mt}{{\ifmmode m_{t}    \else $m_{t}$\fi}}
\newcommand{\mts}{{\ifmmode m_{t}^2    \else $m_{t}^2$\fi}}

\newcommand{\mtau}{{\ifmmode m_{\tau}  \else $m_{\tau}$\fi}}
\newcommand{\dpp}{{\ifmmode \delta_{pert} \else $\delta_{pert}$\fi}}
\newcommand{\dnp}{{\ifmmode\delta_{non-pert}\else$\delta_{non-pert}$\fi}}
\newcommand{\dew}{{\ifmmode \delta_{\rm EW}\else $\delta_{\rm EW}$\fi}}
\newcommand{\rt}{{\ifmmode R_{\tau}  \else $R_{\tau} $\fi}}
\newcommand{\rz}{{\ifmmode R_{Z}  \else $R_{Z} $\fi}}

\newcommand{\swb}{{\ifmmode \sin^2\theta_{\overline{MS}} \else $\sin^2\theta_{\overline{MS}}$\fi}}
\newcommand{\cwb}{{\ifmmode \cos^2\theta_{\overline{MS}} \else $\cos^2\theta_{\overline{MS}}$\fi}}

\newc\AIPCP[3] {{\em AIP Conf. Proc.} {\bf #1} (#2) #3}
\newc\AJ[3] {{\em Astrophys. J.} {\bf #1} (#2) #3}
\newc\AMS[3] {{\em Ann. Math. Statist.} {\bf #1} (#2) #3}
\newc\AP[3] {{\em Ann. Phys.} {\bf #1} (#2) #3}
\newc\APJ[3] {{\em Astropart. J.} {\bf #1} (#2) #3}
\newc\APP[3] {{\em Astropart. Phys.} {\bf #1} (#2) #3}
\newc\APS[3] {{\em Astrophys. J. Suppl.} {\bf #1} (#2) #3}
\newc\ARNPS[3] {{\em Ann. Rev. Nucl. Part. Sci.} {\bf C#1} (#2) #3}
\newc\BA[3] {{\em Bayesian Anal.} {\bf C#1} (#2) #3}
\newc\CPC[3] {{\em Comput. Phys. Commun.} {\bf C#1} (#2) #3}
\newc\CP[3] {{\em Contemp. Phys.} {\bf #1} (#2) #3}
\newc\EPJ[3] {{\em Euro. Phys. Journ.} {\bf C#1} (#2) #3}
\newc\JCAP[3] {{\em JCAP} {\bf #1} (#2) #3}
\newc\JHEP[3] {{\em JHEP} {\bf #1} (#2) #3}
\newc\JPG[3] {{\em J. Phys.} {\bf G #1} (#2) #3}
\newc\IJMP[3] {{\em Int. J. Mod. Phys.} {\bf A #1} (#2) #3}
\newc\MNRAS[3] {{\em Mon. Not. Roy. Astron. Soc.} {\bf #1} (#2) #3}
\newc\MPL[3] {{\em Mod. Phys. Lett.} {\bf A #1} (#2) #3}
\newc\NAR[3] {{\em New Astron. Rev.} {\bf #1} (#2) #3}
\newc\NCA[3] {{\em Nuovo Cimento} {\bf #1} (#2) #3}
\newc\NIM[3] {{\em Nucl. Instrum. Methods} {\bf #1} (#2) #3}
\newc\NIMA[3] {{\em Nucl. Instrum. Methods} {\bf A #1} (#2) #3}
\newc\NAT[3] {{\em Nature} {\bf #1} (#2) #3}
\newc\NPB[3] {{\em Nucl. Phys.} {\bf B #1} (#2) #3}
\newc\NPA[3] {{\em Nucl. Phys.} {\bf A #1} (#2) #3}
\newc\NPPS[3] {{\em Nucl. Phys. Proc. Suppl.} {\bf #1} (#2) #3}
\newc\PLB[3] {{\em Phys. Lett.} {\bf B #1} (#2) #3}
\newc\PR[3] {{\em Phys. Rep.} {\bf #1} (#2) #3}
\newc\PRL[3] {{\em Phys. Rev. Lett.} {\bf #1} (#2) #3}
\newc\PRD[3] {{\em Phys. Rev.} {\bf D #1} (#2) #3}
\newc\PRC[3] {{\em Phys. Rev.} {\bf C #1} (#2) #3}
\newc\PTP[3] {{\em Prog. Theor. Phys.} {\bf #1} (#2) #3}
\newc\RMP[3] {{\em Rev. Mod. Phys.} {\bf #1} (#2) #3 }
\newc\RPP[3] {{\em Rept. Prog. Phys.} {\bf #1} (#2) #3 }
\newc\SC[3] {{\em Science} {\bf #1} (#2) #3 }
\newc\ZPC[3] {{\em Z. Phys.} {\bf C #1} (#2) #3}
\newc\Err[3] {{\em Erratum-ibid.} {\bf #1} (#2) #3 }

\journal{Physics Letters B}

\begin{document}

\begin{frontmatter}


%
\title{Perspectives of direct Detection of supersymmetric Dark Matter in the NMSSM
}

\author[label1]{C. Beskidt}\ead{conny.beskidt@kit.edu}
\author[label1]{W. de Boer}\ead{wim.de.boer@kit.edu}
\author[label1,label2]{D.I. Kazakov}\ead{KazakovD@theor.jinr.ru}
\author[label1]{Stefan Wayand}\ead{stefan.wayand@kit.edu}
\address[label1]{Institut f\"ur Experimentelle Kernphysik, Karlsruhe Institute of Technology, P.O. Box 6980, 76128 Karlsruhe, Germany}
\address[label2]{Bogoliubov Laboratory of Theoretical Physics, Joint Institute for Nuclear Research, 141980, 6 Joliot-Curie, Dubna, Moscow Region, Russia}

\begin{abstract}

In the Next-to-Minimal-Supersymmetric-Standard-Model (NMSSM) the lightest supersymmetric particle (LSP) is a candidate for the dark matter (DM) in the universe. It is a mixture from the various gauginos and Higgsinos and can be bino-, Higgsino- or singlino-dominated. Singlino-dominated LSPs can have very low cross sections below the neutrino background from coherent neutrino scattering which is limiting the sensitivity of future direct DM search experiments. However, previous studies suggested that the combination of both, the spin-dependent (SD) and spin-independent (SI) searches are sensitive in complementary regions of parameter space, so considering both searches  will allow to explore practically the whole parameter space of the NMSSM.
In this letter, the different scenarios are investigated with a new scanning technique, which reveals that significant regions of the NMSSM parameter space cannot be explored, even if one considers both, SI and SD, searches.

\end{abstract}

\end{frontmatter}


\section{Introduction}
\label{Introduction}
Experimental evidence shows that roughly 85\% of the matter in the universe consists of dark matter (DM) \cite{Ade:2015xua}, presumably made at least partially of Weakly Interacting Massive Particles (WIMPs). 
Supersymmetry (SUSY) \cite{Haber:1984rc,deBoer:1994dg,Martin:1997ns,Kazakov:2000ra} can provide a perfect WIMP candidate: the Lightest Supersymmetric Particle (LSP), in many models the lightest neutralino, has all the required WIMP properties: it is neutral, massive, stable and weakly interacting. The observed relic density is inversely proportional to the annihilation cross section \cite{Jungman:1995df} and indeed the LSP  annihilation cross section can  give the right amount of DM in the universe.  This annihilation cross section is required to be some 10 orders of magnitude higher than the limits on the scattering cross section between WIMPs and nuclei, as found in the direct DM detection experiments, which try to detect WIMPs by measuring the recoil of a DM particle off a nucleus in deep underground experiments, see e.g. Refs. \cite{Undagoitia:2015gya,Mayet:2016zxu}.  These many orders of magnitude between the scattering and annihilation cross section are easily explained in SUSY by a combination of the exchanged particle being a Higgs boson, which hardly couples to a nucleus because of the preponderance of light quarks inside a nucleus and the different kinematics from scattering and annihilation.   The direct scattering can  either be proportional to the spin (spin-dependent (SD)) or the scattering is coherent on the whole nucleus, in which case the cross section is enhanced by  the  square of the number of nuclei  of the target material and independent of the spin (spin-independent (SI)).

In the Minimal-Supersymmetric-Standard-Model (MSSM) the LSP is a mixture of gauginos and Higgsinos, with the bino admixture typically being dominant. In this case the present limit of the SI cross section of $2\cdot10^{-10}$ pb  from the LUX 2016 experiment starts to eliminate a significant fraction of the parameter space \cite{Akerib:2013tjd,Akerib:2016lao}.  Limits on the SD cross section are weaker and therefore neglected in the MSSM. With future expected sensitivity on the SI cross section of $10^{-13}$ pb \cite{Aalbers:2016jon} almost the whole parameter space will be accessible in the MSSM, so one would expect to either discover WIMP scattering or exclude the MSSM as the origin of DM.

 However, in the Next-to-Minimal-Supersymmetric-Standard-Model (NMSSM) the situation is different, since the introduction of a Higgs singlet leads to an additional singlino. The Higgs singlet allows to avoid heavy stop masses and avoids the so-called $\mu - problem$, see e.g. \cite{Ellwanger:2009dp}. The LSP will mix with the singlino as well. So the LSP can become predominantly bino-, Higgsino- or singlino-like or be a mixture of them. The larger diversity of the LSP properties has led to many  studies of direct DM detection in the NMSSM, see e.g. \cite{Cerdeno:2004xw,Gunion:2005rw,Cerdeno:2007sn,Hugonie:2007vd,Belanger:2008nt,Das:2010ww,
Barenboim:2011fs,AlbornozVasquez:2012px,Kozaczuk:2013spa,Ellwanger:2014dfa,
Badziak:2015exr,Xiang:2016ndq,Cao:2016cnv,Bednyakov:1999yr,Bednyakov:1998ie,Bednyakov:1998is,Ellwanger:2016sur}.

If  the LSP is predominantly a singlino, it may hardly couple to any SM particle. In this case the non-observation of WIMP scattering may not exclude the NMSSM as the origin of DM, as was studied before in Ref. \cite{Ellwanger:2014dfa}. Here only the SI limits have been taken into account. 

However, recently SD limits have become available \cite{Akerib:2016lao,Amole:2016pye}, which have raised excitement, since they appeared to be complementary in that they exclude different regions of parameter space and it was suggested that in future the combination of SD and SI searches might be able to explore a large fraction of the  NMSSM parameter space \cite{Xiang:2016ndq,Cao:2016cnv}. 

However, these papers relied on Markov Chain or random  sampling of the NMSSM parameter space, in which case it is difficult to sample all regions of a multidimensional parameter space with highly correlated parameters \cite{Trotta:2008qt,Trotta:2011av}.  
The reason is simple: if 3 parameters are positively correlated, stepping through the parameter space with parameter 1 in one direction, one finds maximum likelihoods fastest, if the next steps of the other two parameters are in the same direction. In the constrained MSSM (NMSSM) the dimensionality of the parameter space is 5(9); in unconstrained models significantly larger.   Without knowing the features of a likelihood function with its typical narrow features from correlated parameters, it is difficult to assure  a complete sampling of the parameter space,
 as was demonstrated before for the 5-D parameter space of the MSSM \cite{deAustri:2006jwj,Feroz:2011bj,Roszkowski:2009ye,Beskidt:2012sk} and the 10-D parameter space of the determination of the cosmological parameters of the CMB background \cite{Slosar:2003da,Mueller:2004se,Trotta:2008qt}.
 
 We therefore use a new sampling technique  assuring that  no  regions of parameter space will be missed in the sampling. The main idea is to project the highly correlated parameter space of the couplings onto a space spanned by uncorrelated Higgs masses, which is only 3-D, if one considers one Higgs boson mass fixed to the measured 125 GeV and the heavy Higgs masses to be degenerate. In this space the couplings are marginalized over by a fit. Hence, the Higgs parameter space is reduced from 7-D to 3-D with largely uncorrelated parameters, which allows for an efficient sampling. An alternative way of explaining the sampling technique is as follows:
 suppose the LHC would have discovered all 7 Higgs bosons of the NMSSM. Would we be able to determine all couplings in the Higgs sector? The answer is: there is not a unique solution, but there are two preferred regions in the parameter space, which we called Scenario I and Scenario II in Ref. \cite{Beskidt:2016egy}. By repeating the fit to determine the couplings for each combination of Higgs boson masses in a 3-D grid of Higgs masses one can delineate the parameter regions of Scenario I and Scenario II.

It is the purpose of this letter to check if there are regions in the NMSSM parameter space, which evade exploration by a combination of SD and SI searches.   
We find that there are indeed regions of parameter space, which have cross sections below the ``neutrino floor'', both for the SD and SI searches.  Below the ``neutrino floor'' direct detection will be difficult, because of the high background from  the coherent scattering of  neutrinos, which cannot be shielded in DM experiments. Only tails in the recoil spectrum, annual modulation or directional dependence of the events might allow to separate WIMP scattering from neutrino backgrounds given enough statistics, see Ref. \cite{OHare:2016pjy} and references therein.  Since in parameters regions near or below the neutrino floor the LSP is almost a pure singlino, these regions are not accessible at the LHC either. 

After a short summary of the neutralino sector in the NMSSM and the elastic scattering processes, we discuss the fit strategy. We conclude by summarizing the impact of the DM constraints from future experiments on the NMSSM parameter space.

\section{Semi-constrained NMSSM}
\label{nmssm}

Within the NMSSM the Higgs fields consist of the two Higgs doublets ($H_u, H_d$), which appear in the MSSM as well, but the NMSSM has
an additional complex Higgs singlet $S$. 
The addition of a Higgs singlet yields more parameters in the Higgs sector to cope with the interactions between the singlet and the doublets and the singlet self interaction. 

In the following we restrict the parameter space by assuming unification of couplings and masses at the GUT scale of about $2 \cdot 10^{16}$ GeV. Although this restricts the parameter space, it is a well motivated region of para\-meter space and it will be interesting to see if this region is within reach of the future experiments. In this case we have the GUT scale parameters of the Constrained-Minimal-Supersymmetric-Standard-Model (CMSSM): $\mzero$ and $\mhalf$, where $\mzero$($\mhalf$) are the common mass scales at the GUT scale of the spin 0(1/2) SUSY particles,  the trilinear coupling $A_0$  of the CMSSM Higgs sector and $\tan\beta$, the ratio of vacuum expectation values (vev) of the neutral components of the SU(2) Higgs doublets, i.e. $\tan\beta\equiv v_u/v_d$. For the NMSSM one has to add the coupling $ \lambda$ between the singlet and the doublets from the term $\lambda S H_u\cdot  H_d$ and  $\kappa$, the self-coupling of the singlet from the term $\kappa S^3/3$; $A_\lambda$ and $A_\kappa$ are the corresponding trilinear soft breaking terms; 
$\mu_{eff}$ represents an effective Higgs mixing parameter.

 So in total the semi-constrained NMSSM has nine free parameters:
\begin{equation}
 \mzero,~ \mhalf,~A_0,~ \tan\beta,  ~ \lambda, ~\kappa,  ~A_\lambda, ~A_\kappa, ~\mu_{eff}.
\label{params}
\end{equation}

The effective Higgs mixing parameter  is related to the vev of the singlet $s$ via the coupling $\lambda$, i.e. $\mu_{eff}\equiv\lambda s$. Being proportional to a vev,  $\mu_{eff}$ is naturally of the order of the electroweak scale, thus avoiding the $\mu$-problem \cite{Ellwanger:2009dp}. 
The supersymmetric partner of the singlet leads to an additional Higgs\-ino, thus extending the neutralino sector from 4 to 5 neutralinos. This leads to modifications  of the SI and SD cross sections, which are discussed in the following subsections.

\subsection{The NMSSM neutralino sector}
\label{neu-char-sector}

Within the NMSSM the singlino, the superpartner of the Higgs singlet, mixes with the gauginos and Higgsinos, leading to an additional fifth neutralino. The resulting mixing matrix reads \cite{Ellwanger:2009dp,Staub:2010ty}:

\beq\label{eq1}
{\cal M}_0 =
\left( \ba{ccccc}
M_1 & 0 & -\frac{g_1 v_d}{\sqrt{2}} & \frac{g_1 v_u}{\sqrt{2}} & 0 \\
0 & M_2 & \frac{g_2 v_d}{\sqrt{2}} & -\frac{g_2 v_u}{\sqrt{2}} & 0 \\
-\frac{g_1 v_d}{\sqrt{2}} & \frac{g_2 v_d}{\sqrt{2}} & 0 & -\mu_\mathrm{eff} & -\lambda v_u \\
\frac{g_1 v_u}{\sqrt{2}} & -\frac{g_2 v_u}{\sqrt{2}}& -\mu_\mathrm{eff}& 0 & -\lambda v_d \\
0& 0& -\lambda v_u&  -\lambda v_d & 2 \kappa s 
\ea \right)
\eeq
with the gaugino masses $M_1$, $M_2$, the gauge couplings $g_1$, $g_2$ and the Higgs mixing parameter $\mu_{eff}$ as parameters.
Furthermore, the vacuum expectation values of the two Higgs doublets $v_d$,$v_u$, the singlet $s$ and the Higgs couplings $\lambda$ and $\kappa$ enter the neutralino mass matrix. 
 
The upper left $4 \times 4$ submatrix of the neutralino mixing matrix corresponds to the MSSM neutralino mass matrix, see e.g. Ref. \cite{Martin:1997ns}. 

The neutralino mass eigenstates are obtained from the diagonalization of ${\cal M}_0$ in Eq. \ref{eq1} and are linear combinations of the gaugino and Higgsino states: 
\beq\label{eq5}
\footnotesize{
\tilde{\chi}^0_i = {\cal N}(i,1) \left | \tilde{B} \right\rangle +{\cal N}(i,2) \left | \tilde{W}^0 \right\rangle 
  +{\cal N}(i,3) \left | \tilde{H}^0_u \right\rangle +{\cal N}(i,4) \left | \tilde{H}^0_d \right\rangle+{\cal N}(i,5) \left | \tilde{S} \right\rangle.}
\eeq

Typically, the diagonal elements in Eq. \ref{eq1} dominate over the off-diagonal terms, so the neutralino masses are of the order of $M_1$, $M_2$, the Higgs mixing parameter $\mu_{eff}$ for the Higgsinos and in case of the NMSSM $2 \kappa s \ = \ 2 (\kappa/ \lambda) \mu_{eff}$ for the singlino-like neutralino. 

 The mass spectrum at the low mass SUSY scales is calculated from the GUT scale input parameters via the renormalization group equations (RGEs), which results in correlated masses including the large radiative corrections from the GUT scale to the electroweak scale. The gaugino masses at the electroweak scale are proportional to $m_{1/2}$ \cite{Haber:1984rc,deBoer:1994dg,Martin:1997ns,Kazakov:2015ipa}:
\beq\label{eq3}
M_1\approx 0.4 m_{1/2},~ 
M_2\approx 0.8 m_{1/2},~
M_3\approx M_{\tilde{g}} \approx 2.7 m_{1/2}.  
\eeq 

In the CMSSM the Higgs mixing parameter $\mu$ is typically much larger than $m_{1/2}$ to fulfill radiative electroweak symmetry breaking (EWSB) \cite{Haber:1984rc,deBoer:1994dg,Martin:1997ns,Kazakov:2015ipa}, which leads to a bino-like lightest neutralino.
In the NMSSM $\mu_{eff}$ is an input parameter, which is naturally  of the order of the electroweak scale.
In such natural NMSSM scenarios the lightest neutralino is singlino- or Higgsino-like and its mass can be degenerate with the second and third neutralino, all of which have a mass of the order of $\mu_{eff}$. Bino-like neutralinos are also possible within the NMSSM but they require large values of $\mu_{eff} >> M_1$. This is not excluded, but not expected in natural NMSSM models. However, if the LSP in the NMSSM is bino-like, the situation is similar to the MSSM, which has been studied in great detail previously \cite{Beskidt:2014oea}. So in this letter we will concentrate on LSPs being singlino- or Higgsino-like in the NMSSM. 

The amount of the Higgsino and singlino content of the lightest neutralino depends on the ratio and the absolute value of $\kappa$ and $\lambda$, as can be seen from the coefficient ${\cal M}_0(5,5)=2 \kappa s \ = \ 2 (\kappa/ \lambda) \mu_{eff} $. 
The Higgsino fraction, which determines the coupling to the Higgs, is crucial for the elastic scattering cross section, since this proceeds mainly via the exchange of a Higgs boson.

\subsection{Elastic WIMP-Nucleon Scattering}
\label{chap4:sec6}

 A WIMP might be detected  by measuring the recoil of  a nucleus after an elastic scattering of a WIMP on a nucleus taking place. Since such collisions are non-relativistic, only two cases need to be considered \cite{Goodman:1984dc}: the spin-spin interaction (SD), where the WIMP couples to the spin of the nucleus, and the scalar interaction (SI), where the WIMP couples to the mass of the nucleus.

 The SI cross section is proportional to the Higgsino content of the lightest neutralino $\sigma_{SI} \propto N_{13}^2 + N_{14}^2$ and to the mass of the nucleus squared, which leads to a substantial enhancement for heavy nuclei \cite{Drees:1993bu}. 
In addition, the cross section includes the effective quark form-factors which are similar for protons and neutrons and increase for large values of the strange quark content. However, the  quark form-factors derived from pion-nucleon scattering measurements suffer from large uncertainties \cite{Ellis:2008hf}. In addition, these measurements deviate from the form-factors resulting from lattice calculations. We calculate all DM cross sections with micrOMEGAs 3.6.9.2 \cite{Belanger:2013oya}. The default form-factors given in micrOMEGAs are taken from the average of a variety of different measurements and lattice calculations \cite{Belanger:2010pz}. The extreme values for the form factors lead to variations in the predicted cross section of about 20\%.  

The experimental best limit on the SI WIMP nucleon cross section is given by the LUX experiment \cite{Szydagis:2016few}. It excludes discovery claims by DAMA/LIBRA \cite{Bernabei:2010mq} and CoGeNT \cite{Aalseth:2012if}.
The SI cross section is inversely proportional to the Higgs mass squared, so the prediction of two light scalar Higgs bosons can enhance the SI cross section in the NMSSM. However, a negative interference between them suppresses the SI cross section if the two lightest Higgs bosons are close in mass. In this case the predicted cross section is below the current LUX limits, which has been discussed in more detail in Refs. \cite{Beskidt:2014oea,Badziak:2015exr,Cao:2016cnv}.
However, the SD cross section, which proceeds mainly by $Z^0$ exchange, does not suffer from such "blind" spots, which have a steep probability distribution in the parameter space, as demonstrated in Fig. 4c from Ref. \cite{Beskidt:2014oea}. 

The dominant diagram for the SD scattering is the $Z^0$ boson exchange. The corresponding cross section includes the difference of the Higgsino components $\sigma_{SD} \propto |N_{13}^2 - N_{14}^2 |$. If the admixture of the two Higgsino components are large but similar, the SD cross section can become small. 
But then the SI cross section ($\propto N_{13}^2 + N_{14}^2$) will be large, so they do not become small simultaneously. The calculation of the nuclear matrix elements is at zero momentum transfer equivalent to the calculation of the average spins for neutrons and protons, while the corresponding coefficients can be extracted from data on polarized deep inelastic scattering \cite{Jungman:1995df}. Uncertainties in the experimental determination of these coefficients lead to variations in the predicted rates for WIMP detection as already discussed above for the SI cross section. 
The current best limit on the SD cross section is given by LUX for the WIMP-neutron interaction \cite{Akerib:2016lao}, as the majority of the nuclear spin is carried by the unpaired neutron in the Xenon isotopes. PICO-2L gives the best limit on the SD WIMP-proton cross section \cite{Amole:2016pye} because of the single unpaired proton in $C_3 F_8$ providing a better sensitivity for SD WIMP-proton interactions. 
Naively, one would expect the SD cross sections to be the same for neutrons and protons. However, they are different because of the proton and neutron form factors which leads to $\sigma_n \approx 0.77 \sigma_p$ \cite{Badziak:2015exr}. If  we take these different cross sections for proton and neutrons into account, the LUX experiment is more sensitive than the PICO experiment, so we continue to consider SD neutron cross sections, thus following Ref. \cite{Cao:2016cnv}.  

The experimental limits require values for the exposure, which depends on the local DM density, which takes values between 0.3 and 1.3 $\mathrm{GeV/cm^3}$, see e.g. \cite{Weber:2009pt}. This uncertainty leads to a variation of the limit of about a factor of 4. The limits given by different experiments are calculated for a  local DM density of 0.3 $\mathrm{GeV/cm^3}$, which leads to the most conservative limit.

\section{Analysis}
\label{analysis}

The additional particles and their interactions within the NMSSM lead to a large parameter space, even in the well-motivated subspace with unified masses and couplings at the GUT scale. We focus on the semi-constrained NMSSM  and  use the corresponding code NMSSMTools 4.6.0 \cite{Ellwanger:2004xm,Ellwanger:2005dv} to calculate the SUSY mass spectrum from the NMSSM parameters. The Higgs masses depend on radiative corrections, which are calculated using the option 8-2 in NMSSMTools, which means that the full one loop and the full two loop corrections from top and bottom Yukawa couplings are taken into account. NMSSMTools has an interface to micrOMEGAs \cite{Belanger:2013oya}, which was used to calculate the relic density and LSP scattering cross sections.

As discussed in the introduction, we use a systematic sampling technique by considering a space spanned by the masses of the 3 scalar and 2 pseudo-scalar neutral Higgs boson masses $m_{H_i}$ and $m_{A_i}$, as well as the two charged Higgs bosons $m_{H^\pm}$. 
This space reduces to a 3-D parameter space, if one requires $m_{H1}$ or $m_{H2} \approx 125$ GeV with SM couplings and $m_{H3} \approx m_{A2} \approx m_{H^\pm}$ for $M_A >> M_Z$. We took the lightest[second-lightest] and heaviest neutral scalar Higgs boson masses and the lightest pseudo-scalar neutral Higgs mass as remaining masses, so after choosing these three "free" masses all Higgs masses are fixed. 
The "free" masses are distributed over a grid $m_{H1[H2]}$ vs. $m_{H3}$ for different steps in $m_{A1}$ \cite{Beskidt:2016egy}.
These grid boundaries were chosen to lay between
\begin{align}\label{eq4}
5[125]~GeV &< m_{H1[H2]} < 125[500]~GeV \nonumber \\ 
100~GeV &< m_{H3} < 2~TeV \\
5~GeV &< m_{A1} < 500~GeV. \nonumber
\end{align}

For each mass combination the allowed couplings were determined from a fit
which minimizes the following $\chi^2$ function with the parameters of Eq. \ref{params} as free parameters:
\beq\label{eq5}
\chi^2_{tot}=\chi^2_{H_S}+\chi^2_{H_{SM}}+\chi^2_{H_3}+\chi^2_{LEP}.
\eeq 

The $\chi^2$ contributions are \cite{Beskidt:2016egy}

\begin{itemize} 
\item $\chi^2_{H_S}=(m_{H_i} - m_{grid,H_i})^2/\sigma^2_{H_i}$: since one of the light Higgs bosons represent the observed SM Higgs, the other light Higgs boson ${H_i}$ with $i=1,2$ has to be singlet-like. The term $\chi^2_{H_S}$ requires the NMSSM parameters to be adjusted such that the mass of the singlet-like light Higgs boson mass $m_{H_i}$ with $i=1,2$ agrees with the chosen point in the 3-D mass space $m_{grid,H_i}$. 
The value of $\sigma^2_{H_i}$ is set to 2 GeV. 
\item $\chi^2_{H_{SM}}=(m_{H_i} - m_{obs})^2/\sigma^2_{SM}+\sum_i (c^i_{H_i} - c_{obs})^2/\sigma^2_{coup}$: the other light Higgs boson ${H_i}$ with $i=1,2$ has to represent the observed Higgs boson with couplings close to the SM couplings, as required by the last term.  $c^i_{H_i}$ represents the reduced couplings of $H_i$  which is the ratio of the coupling of $H_i$ to particle $i=f_u,f_d,W/Z,\gamma$ divided by the SM coupling. The observed couplings $c_{obs}$ agree within 10\% with the SM couplings, so $\sigma^2_{coup}=0.1$. The first term is analogous to the term for  $m_{H_S}$, except that the mass of the second light Higgs boson should have the observed Higgs boson mass, so $m_{obs}$ is set to $125.4$ GeV. The corresponding uncertainty $\sigma^2_{SM}$ equals $1.9$ GeV and results from the linear addition of the experimental and theoretical ($1.5$ GeV) uncertainties. 
\item $\chi^2_{H_3}=(m_{H_3} - m_{grid,H_3})^2/\sigma^2_{H_3}$: as $\chi^2_{H_S}$, but for the heavy scalar Higgs boson $H_3$.
\item $\chi^2_{LEP}$: includes the LEP constraints on the couplings of a light Higgs boson below 115 GeV and the limit on the chargino mass as discussed in Ref. \cite{Beskidt:2014kon}.      
\end{itemize}

The $\chi^2$ function is insensitive to the SUSY mass parameters $m_0$ and $m_{1/2}$, so these were fixed to 1 TeV. This mass point leads to a sparticle spectrum consistent with the current limits of the direct SUSY searches from the LHC \cite{ATLAS-CONF-2016-078,CMS-PAS-SUS-16-014}. In addition to the LHC constraints, further constraints, like constraints from B-physics, are calculated and checked within NMSSMTools. The fit finds the best values of the couplings, but there is no unique solution, as can be seen already from the approximate expression for the Higgs mass  
\cite{Ellwanger:2009dp}: 
\beq\label{eq4}
M_{H}^2\approx M_Z^2\cos^2 2\beta+ \Delta_{\tilde{t}} + \lambda^2 v^2 \sin^2 2\beta - \frac{\lambda^2}{\kappa^2}(\lambda-\kappa \sin 2 \beta)^2.  
\eeq 

The first two terms are identical to the expression in the CMSSM, where the first tree level term can become as large as  $M_Z^2$ for large $\tan\beta$, but in the CMSSM the difference between  $M_Z$  and 125 GeV  has to originate mainly from the logarithmic stop mass correction $\Delta_{\tilde{t}}$ \cite{Ellis:1990nz}. 
The two remaining terms originate from the mixing with the singlet of the NMSSM.
The SM Higgs boson in the NMSSM is fulfilled within two regions of the parameter space, as was determined in a previous paper \cite{Beskidt:2016egy}. The first region has  large values of $\lambda$ and $\kappa$ and small values of $\tan\beta$ which we call {\it Scenario I}. Here the tree level mass of the Higgs is large due to the mixing with the singlet. Another possibility, which we call {\it Scenario II}, are  small values for $\lambda$ and $\kappa$, which requires large values of $\tan\beta$ in order to reach a Higgs mass of 125 GeV. 
 Within these two scenarios either the lightest or the second lightest Higgs can be the discovered 125 GeV Higgs boson. The range of the couplings allowed by the fit in both scenarios has been given before  \cite{Beskidt:2016egy}.

The specific scenarios have distinctly different features since the range of the couplings differ. However, the ratio of $\lambda$ and $\kappa$, which determines the Higgsino-singlino mixture of the LSP, can be the same in both scenarios, so the singlino content can be the same in both scenarios, as can be seen from the ${\cal M}_0(5,5)$ element in Eq. \ref{eq1}.

\begin{figure}
\begin{center}
\hspace{-2.5cm}
\begin{minipage}{\textwidth}
\begin{tabular}{cc|c}
&{  $\tilde{H}$}& { $\tilde{S}$} \\
&  \multirow{9}{*}{\includegraphics[width=0.44\textwidth]{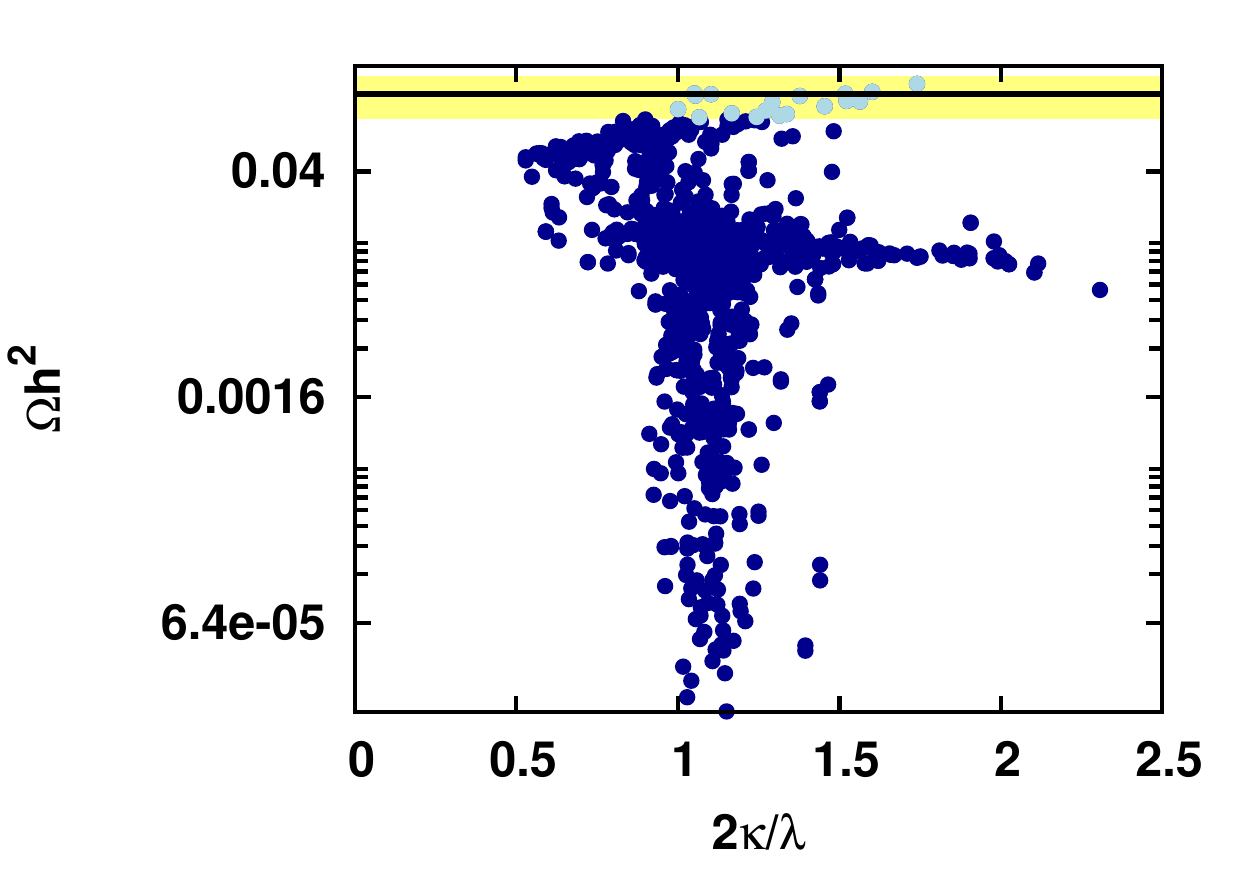}} & \multirow{9}{*}{\includegraphics[width=0.44\textwidth]{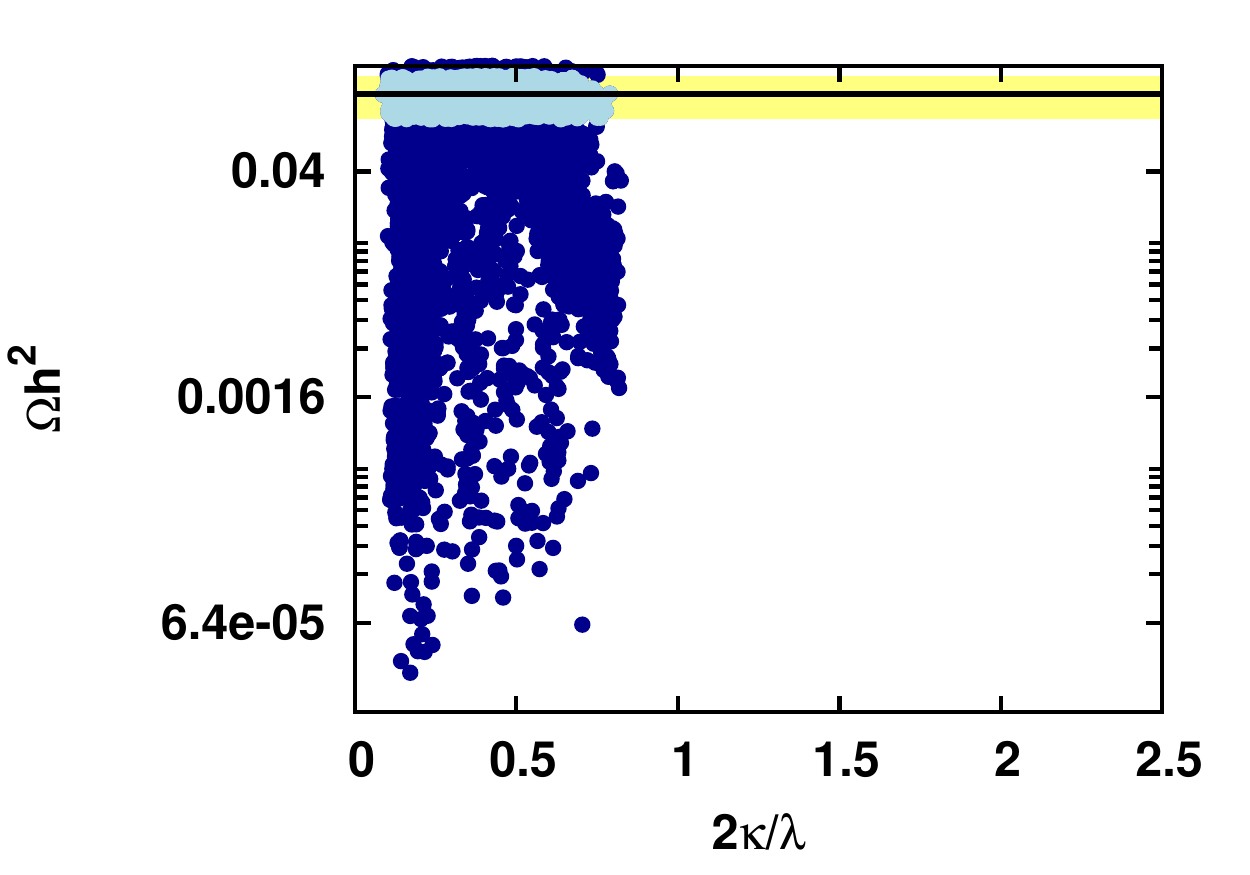}} \\
 &  & \\
 &  & \\
 &  & \\
\footnotesize{Scenario I} & & \\
\scriptsize{(large $\lambda,\kappa$, } &  & \\
\scriptsize{small $\tan\beta$)} &  & \\
 &  & \\
 &  & \\
	\noalign{\smallskip}\cline{2-3}\noalign{\smallskip}
&  \multirow{9}{*}{\includegraphics[width=0.44\textwidth]{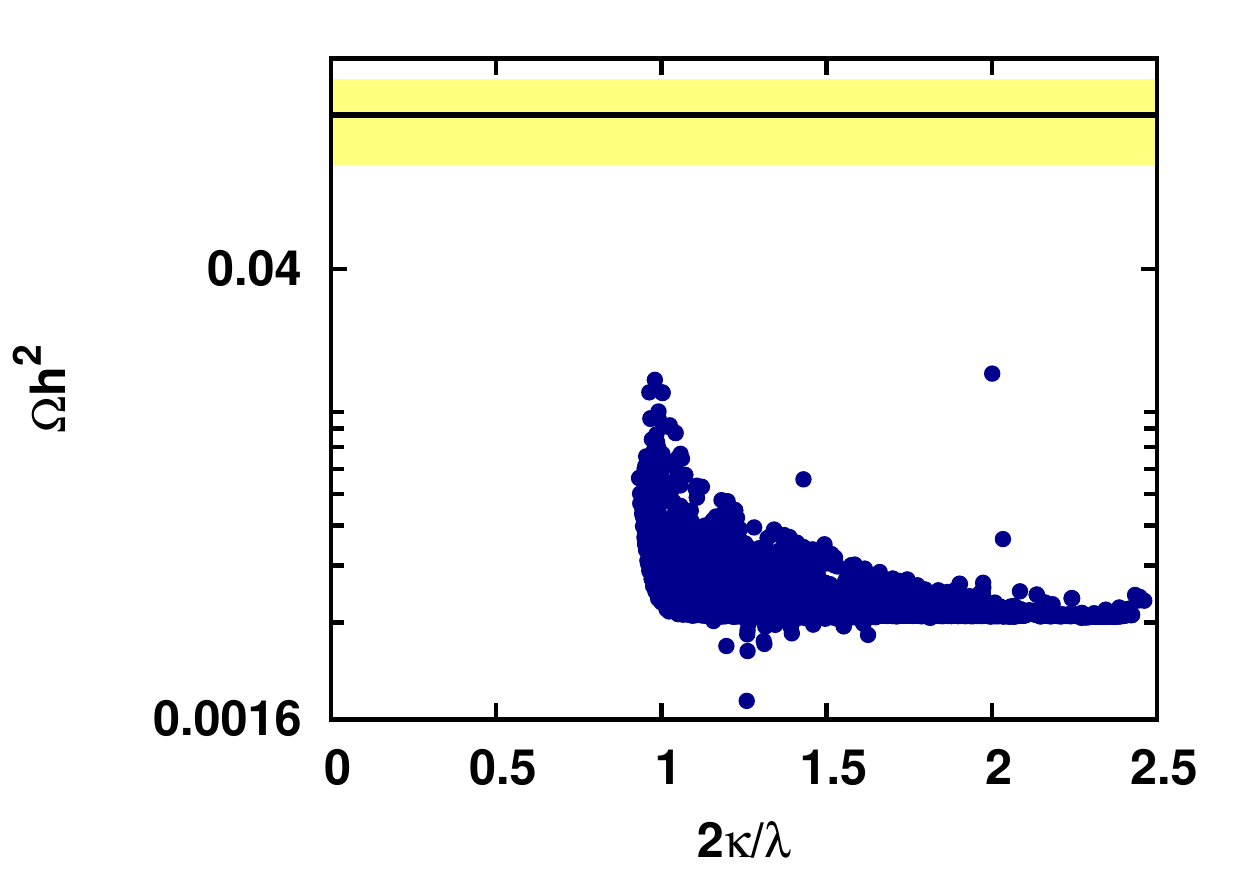}} & \multirow{9}{*}{\includegraphics[width=0.44\textwidth]{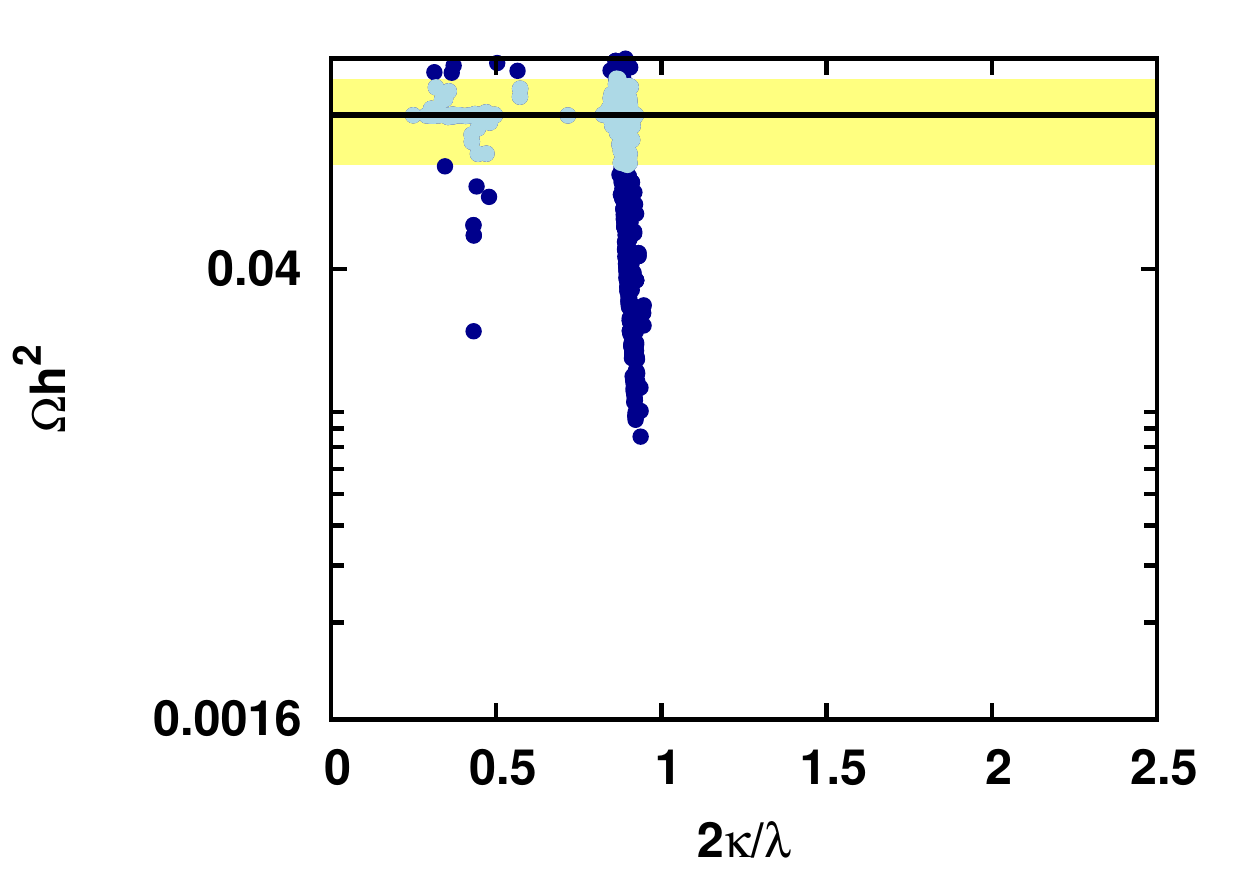}} \\
 &  & \\
 &  & \\
 &  & \\
\footnotesize{Scenario II} & & \\
\scriptsize{(small $\lambda,\kappa$, } &  & \\
\scriptsize{large $\tan\beta$)}  &  & \\
 &  & \\
 &  & \\
\end{tabular}
\end{minipage}
\captionsetup{font=tiny}
\caption[]{Top row:
The relic density  versus $2\kappa/\lambda$ for Scenario I and $m_{H_2}=125$ GeV for the Higgsino (left) and singlino-dominated points (right). Bottom row: as top row, but for  Scenario II. The experimental value of the relic density is represented by the black solid line with the colored band corresponding to the 95\% C.L. region resulting from the linear addition of the experimental and theoretical error. The dark blue points represent Higgsino/singlino-type neutralinos  with a corresponding neutralino content above 0.8. The light blue points saturate the relic density at 95\% C.L.. The correct value of the relic density is easily fulfilled for the singlino-dominated LSPs, while for the Higgsino-type LSPs the predicted relic density is usually below the measured value due to the large coupling to Higgs bosons.}
\label{f3-3}
\end{center}
\end{figure}

Since the Higgsino content is crucial for the SD and SI cross section, we divide the two scenarios further into singlino- and Higgsino-dominated scenarios. This means that either the Higgsino elements $\sqrt{N_{13}^2+N_{14}^2}$ or the singlino element $\sqrt{N_{15}^2}$ are above 0.8. 
All cases can be either fulfilled for the lightest or the second lightest Higgs boson being the SM Higgs boson, which gives in total 8 scenarios (I and II with Higgsino/Singlino LSP and either $m_{H_1}$ or $m_{H_2}$ =125 GeV) be tested against the current SD and SI limits.
Besides the direct DM detection limits the relic density $\Omega h^2$ can be  considered, either as an upper limit, if one assumes other particles contributing to the DM abundance in the universe as well, or one assumes that the LSPs saturate the relic density from the Planck data \cite{Ade:2013zuv}. However, if the relic density is too high, this would over-close the universe and such points are excluded.
For the sampled points the predicted value of the relic density is plotted versus $2\kappa/\lambda$ in Fig. \ref{f3-3}. The dark blue points correspond to the Higgsino- and singlino-dominated points. The top/bottom row shows Scenario I/II.   For the Higgsino-dominated (left) LSPs the relic density is usually below the experimental value because of the large annihilation cross section into $ZZ$ and $W^+W^-$.  In contrast, the singlino-dominated (right) LSP can cover a large range of relic densities,  since many co-annihilation channels can contribute. Co-annihilation is important, because the lightest neutralinos all have similar masses of the order of $\mu_{eff}$.  If co-annihilation is not possible, large relic densities are obtained, because the singlinos hardly couple to SM particles leading to small annihilation cross sections. As mentioned before, such points over-close the universe and are rejected for further analyses.  
The correct relic density can also be fulfilled for resonant annihilation via $Z^0$ or $H$ boson leading to narrow allowed regions around $m_{\tilde{\chi}^1_0} \approx 45$ and $m_{\tilde{\chi}^1_0} \approx 60$ GeV. Resonant annihilation via the light pseudo-scalar Higgs boson $A_1$ is possible for light neutralino masses of the order of a few GeV.

\begin{figure}
\vspace{-0.7cm}
\begin{center}
\hspace{-3.5cm}
\begin{minipage}{\textwidth}
\begin{tabular}{ccc|c}
& &{  $\tilde{H}$}& { $\tilde{S}$} \\
& &  \multirow{7}{*}{\includegraphics[width=0.43\textwidth]{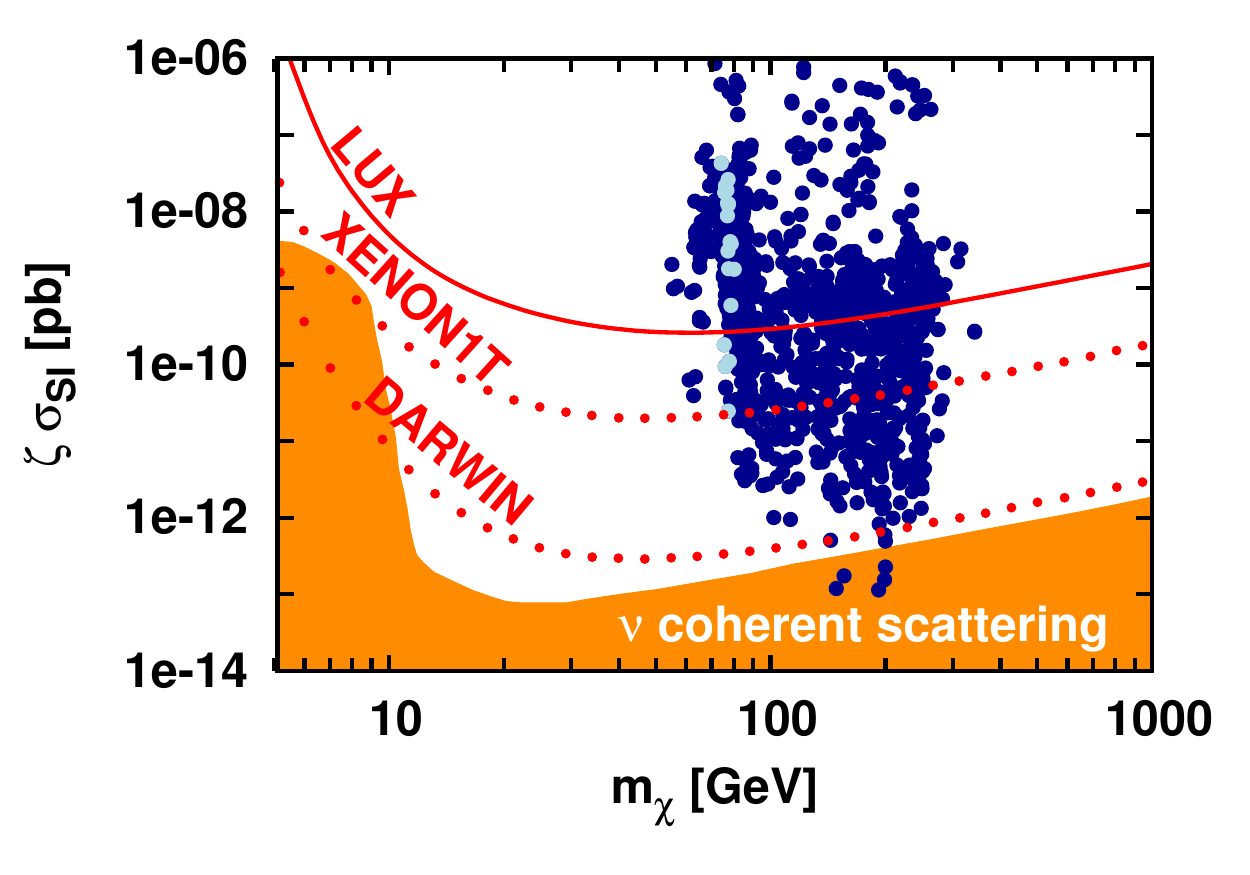}} & \multirow{7}{*}{\includegraphics[width=0.43\textwidth]{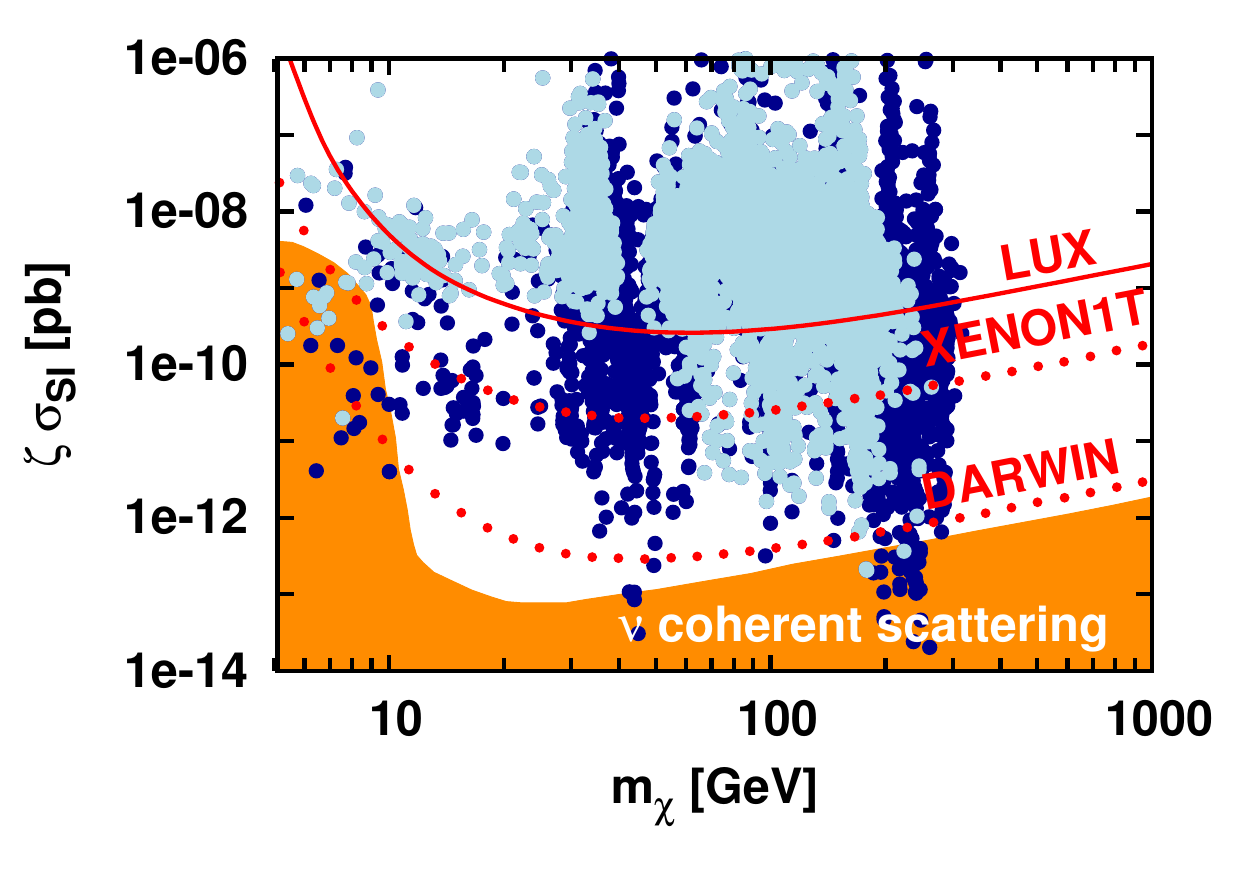}} \\
&  &  & \\
 & &  & \\
& \footnotesize{SI} & & \\
 & &  & \\
 & &  & \\
\footnotesize{\bf{Scenario I} }& &  & \\
\footnotesize{(large $\lambda,\kappa$, }& &  \multirow{9}{*}{\includegraphics[width=0.43\textwidth]{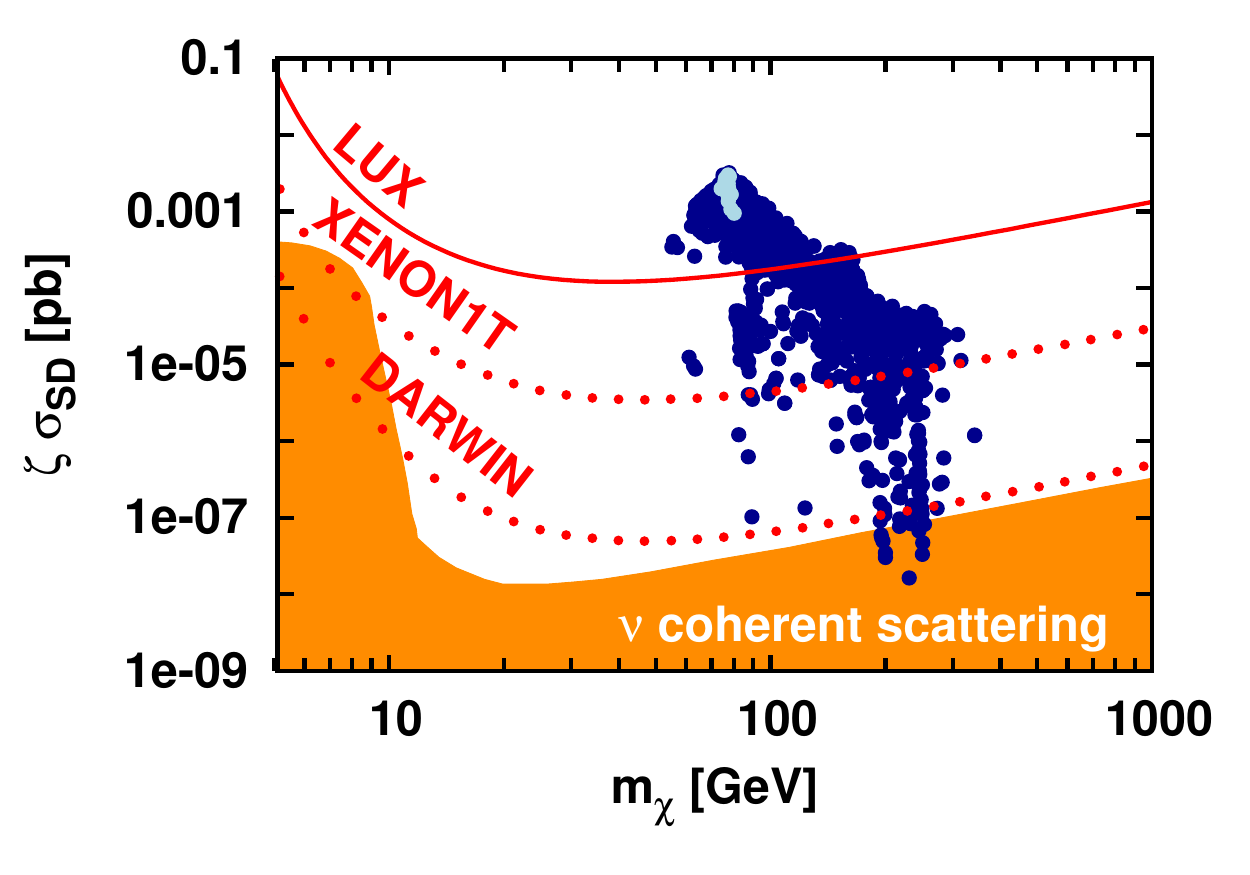}} & \multirow{9}{*}{\includegraphics[width=0.43\textwidth]{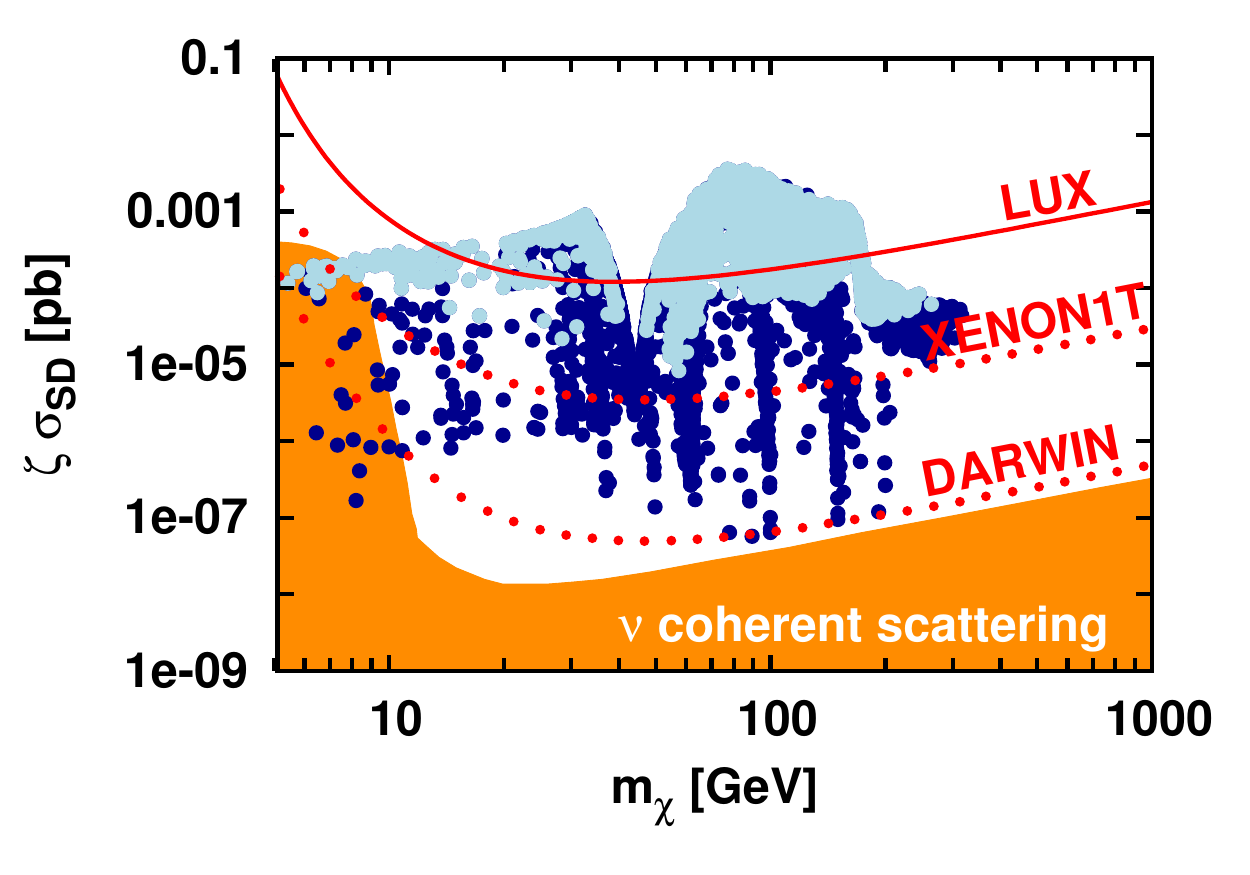}} \\
\footnotesize{small $\tan\beta$)} & &  & \\
 &  & & \\
 &  & & \\
& \footnotesize{SD} & & \\
&  &  & \\
&  &  & \\
\footnotesize{[$m_{H_i}+\zeta \Omega h^2$ }&  &  & \\
\footnotesize{$\zeta <1$\includegraphics[width=0.02\textwidth]{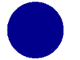}}&  &  & \\
	\noalign{\smallskip}\cline{3-4}\noalign{\smallskip}
\footnotesize{$\zeta=1$\includegraphics[width=0.02\textwidth]{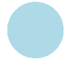}]} & &  \multirow{7}{*}{\includegraphics[width=0.43\textwidth]{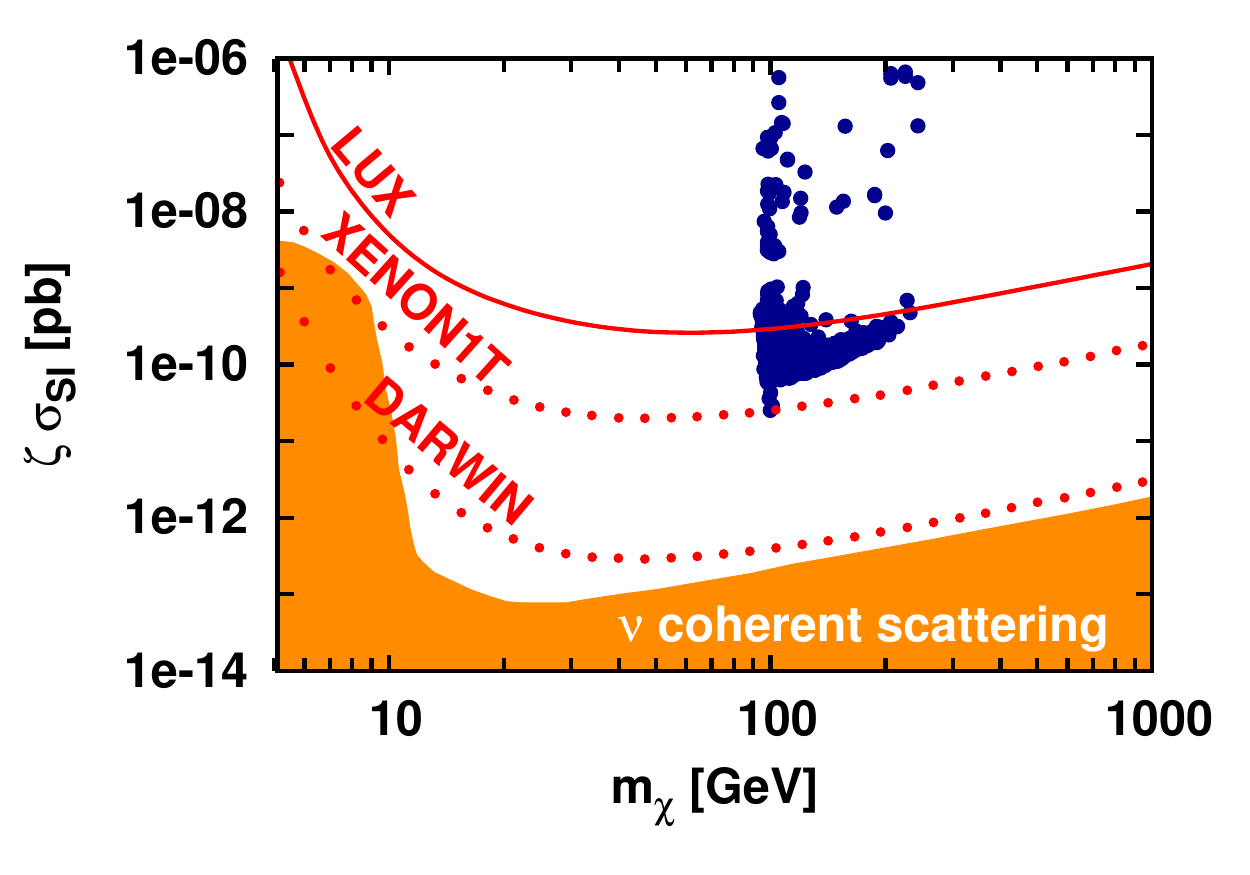}} & \multirow{7}{*}{\includegraphics[width=0.43\textwidth]{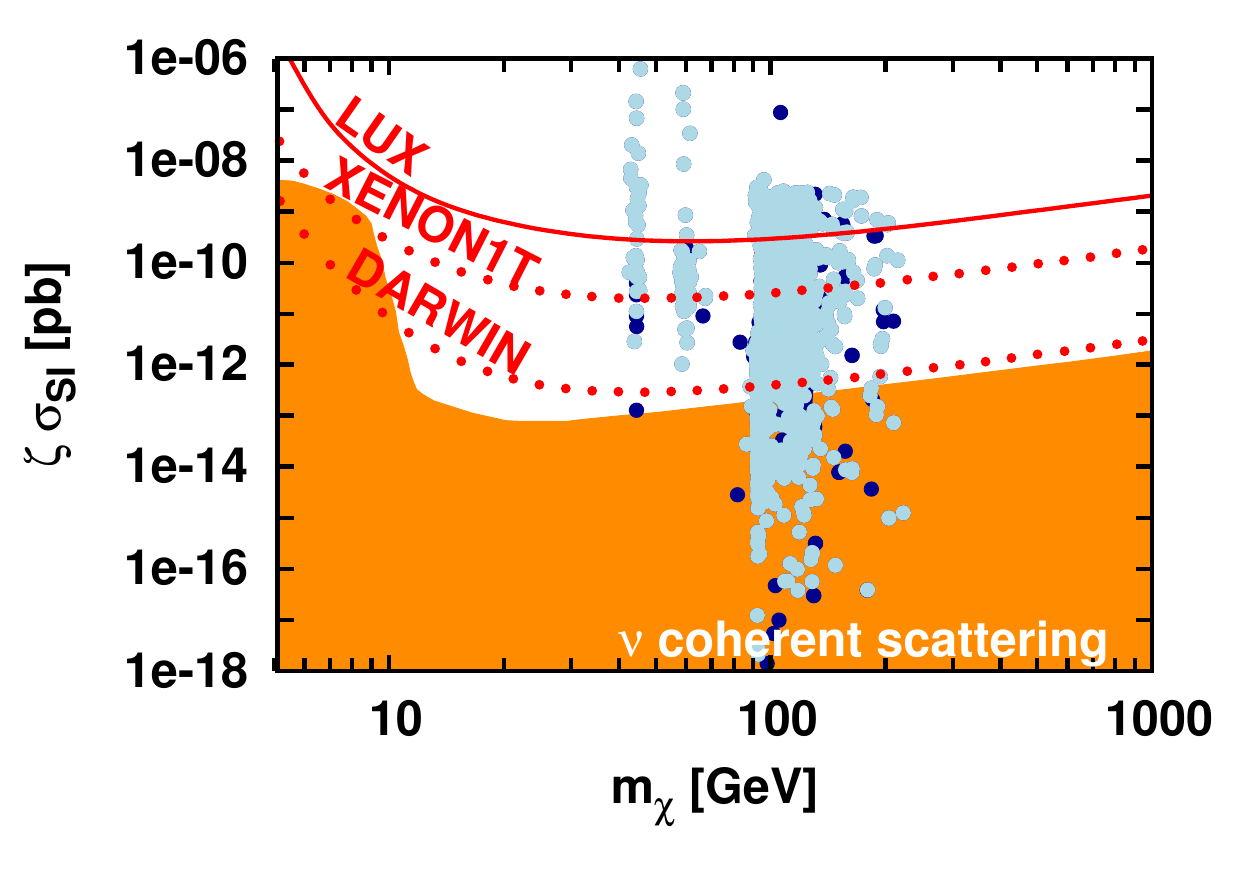}} \\
&  &  & \\
 & &  & \\
& \footnotesize{SI} & & \\
 & &  & \\
 & &  & \\
\footnotesize{\bf{Scenario II} } & &  & \\
\footnotesize{(small $\lambda,\kappa$, }& &  \multirow{9}{*}{\includegraphics[width=0.43\textwidth]{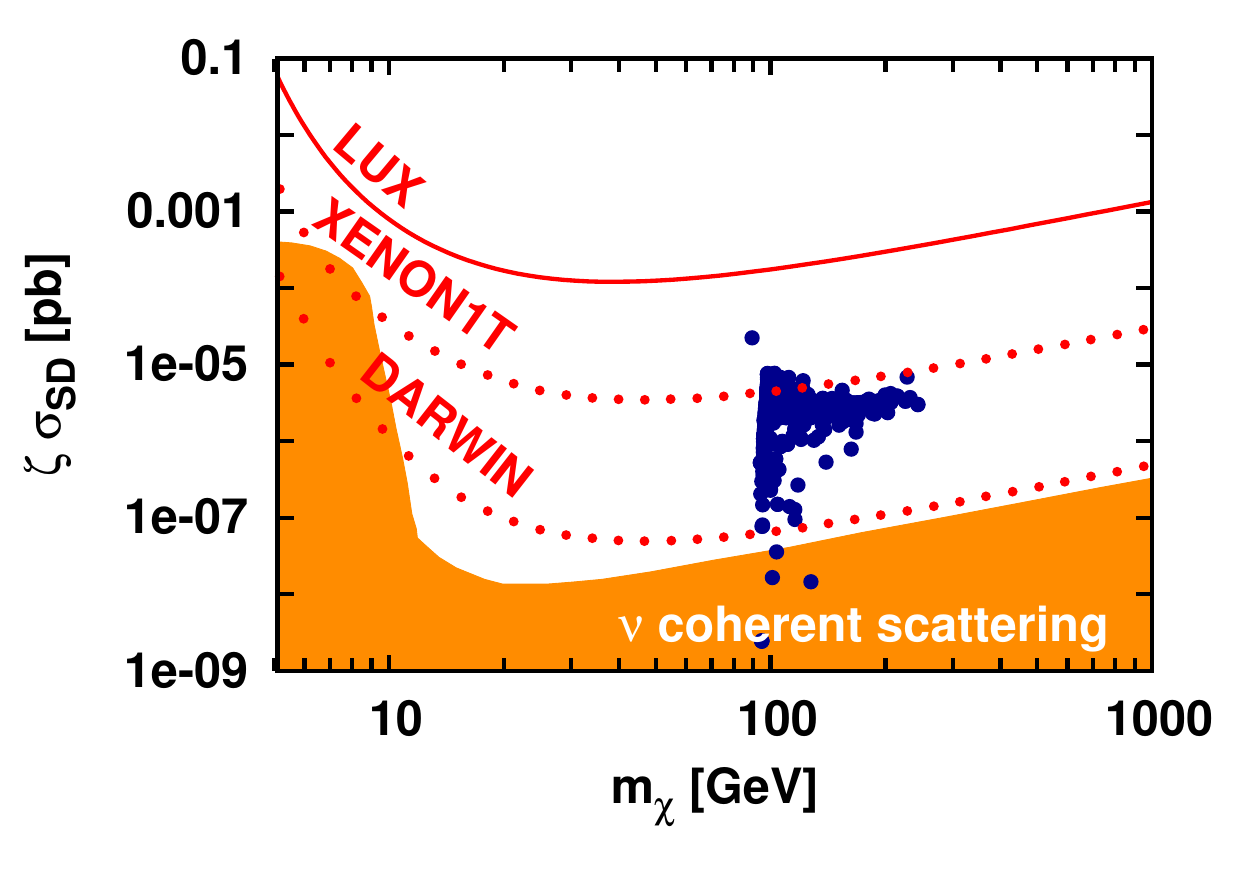}} & \multirow{9}{*}{\includegraphics[width=0.43\textwidth]{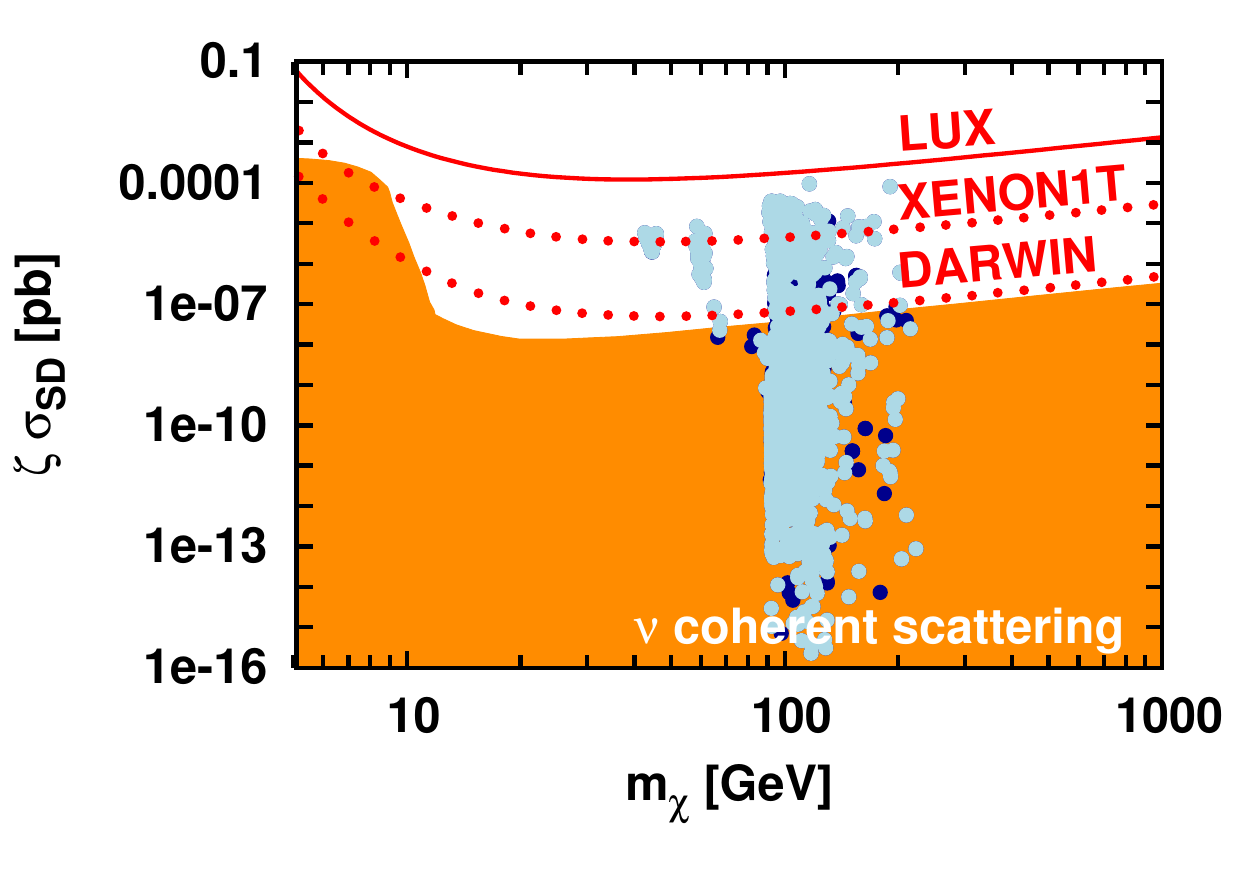}} \\
\footnotesize{large $\tan\beta$) }& &  & \\
 &  & & \\
 &  & & \\
& \footnotesize{SD} & & \\
&  &  & \\
&  &  & \\
&  &  & \\
&  &  & \\
\end{tabular}
\end{minipage}
\captionsetup{font=tiny}
\caption[]{Plots of the reduced scattering cross sections (SI and SD, as indicated) versus the WIMP mass.  The left/right plots represent the Higgsino/singlino-dominated LSPs for Scenario I (top rows) and II (bottom rows). The  Higgs mass $m_{H_2}$ was chosen to be 125 GeV, but the case with  $m_{H_1}=125$ GeV looks similar. The dark blue points fulfill the SM Higgs constraint, while the light blue points also yield the correct relic density. The dark blue points have a cross section multiplied by the sensitivity factor $\zeta=\Omega_{theo}/\Omega_{obs}$. The red solid/dotted lines represent the current/future sensitivities for various experiments. The orange area  is below the neutrino coherent scattering cross section from solar, atmospheric and diffuse supernova neutrinos on nuclei, thus  providing a high background for future DM searches, which makes this region  challenging for  future experiments \cite{OHare:2016pjy}.   }
\label{f4}
\end{center}
\end{figure}

\subsection{Reach of direct DM searches}
The sampled points from the parameter space spanned by the Higgs masses for Scenario I and II with the lowest $\chi^2$ are assumed to be representative for the NMSSM, so these points are compared with the relic density and DM scattering cross section limits. 

As shown in Fig. \ref{f3-3}, many sampled points have an expected relic density $\Omega_{theo}$   below the observed relic density, which is allowed if the DM has additional contributions from other particles, like axions. In this case the sensitivity of direct DM experiments will be reduced by the factor $\zeta=\Omega_{theo}/\Omega_{obs}$. If $\Omega_{theo} > \Omega_{obs}$  the points are excluded,  so $\zeta$ cannot be above 1. In order to calculate the reach of direct DM search experiments we multiply the expected cross section with $min(1, \zeta$) to obtain, what we call the reduced cross section.  
The sampled points can  be projected into the WIMP mass - reduced cross section plane as shown in Fig. \ref{f4} for Scenario I/II for the SI and SD cross sections separately, as indicated. Here the second lightest Higgs boson is the SM Higgs boson. The results for $m_{H1}=125$ GeV are similar.
The left/right plots represent the Higgsino/sing\-lino-dominated LSPs. The dark blue points fulfill the SM Higgs constraint, while the light blue points also yield the correct relic density, which is mostly possible for singlino-dominated LSPs, as shown before in Fig. \ref{f3-3}. For the Higgsino-dominated LSPs the relic density is usually too low. The red solid lines represent the current limits on the SI and SD cross sections, while the red dotted lines are the expectations from the future direct DM experiments XENON1T \cite{Aprile:2015uzo} and DARWIN \cite{Aalbers:2016jon}. The orange area below is the coherent neutrino scattering cross section of solar, atmospheric and diffuse supernova neutrinos on nuclei, which limits the sensitivity of direct detection experiments \cite{Cushman:2013zza}. Points within this area are expected to be challenging to access in the future \cite{OHare:2016pjy}. We choose not to give the percentage of the excluded points, since this number varies strongly with the size of the initial parameter space.

The predicted neutralino mass ranges differ for the different scenarios. For the Higgsino-dominated LSP the mass range starts at around 100 GeV, which is determined by  the lowest value of $\mu_{eff}$ choosen around 100 GeV. For a singlino-dominated LSP the mass can be below $\mu_{eff}$, since the mass is proportional to the ratio of $\kappa$ and $\lambda$. The light neutralino masses in the order of a few GeV results from low values of the lightest pseudo-scalar Higgs boson $A_1$.

\begin{figure}
\begin{center}

\begin{minipage}{\textwidth}
\begin{tabular}{cc}
\footnotesize{\bf Scenario I} \scriptsize{(large $\lambda,\kappa$, small $\tan\beta$)} & \footnotesize{\bf Scenario II} \scriptsize{(small $\lambda,\kappa$, large $\tan\beta$)} \\
 \includegraphics[width=0.49\textwidth]{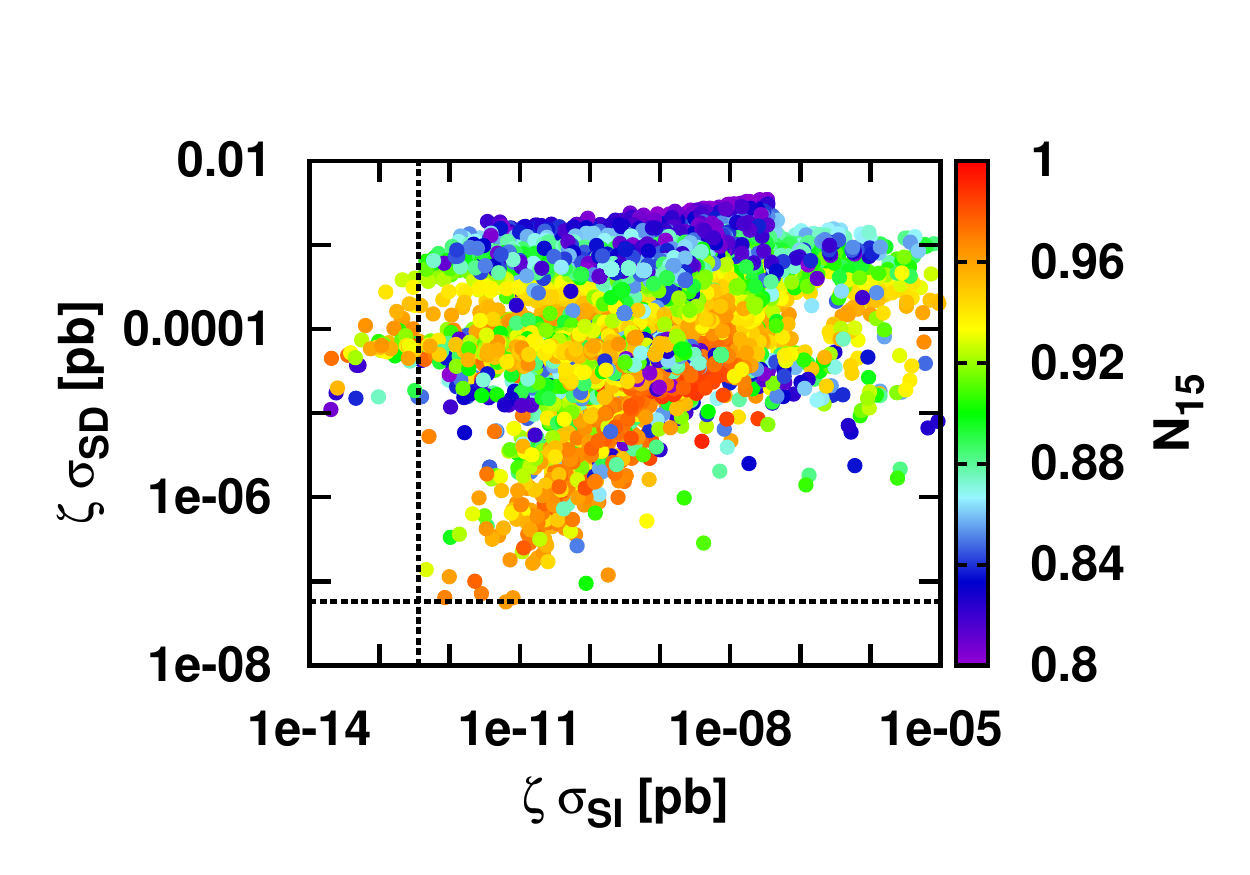}&  \includegraphics[width=0.49\textwidth]{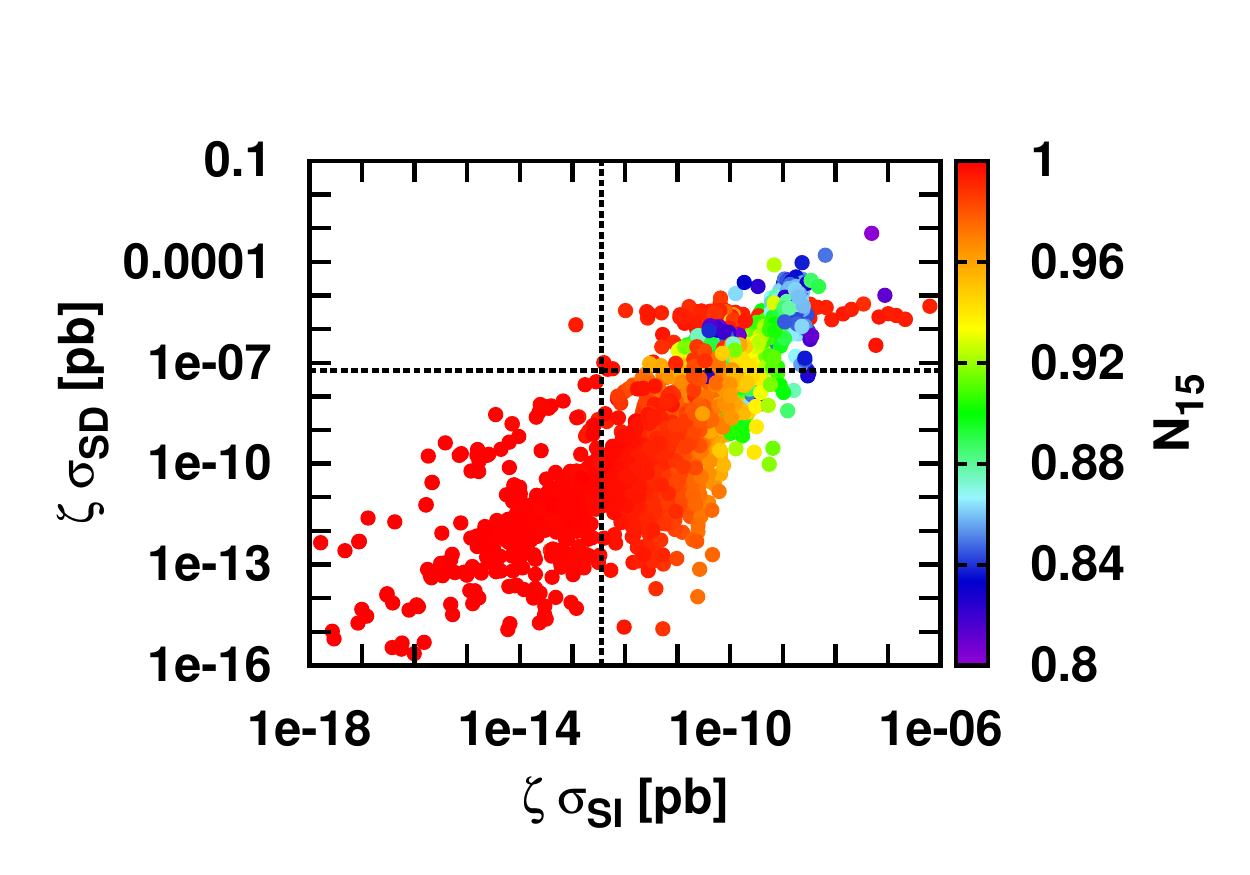} \\
\end{tabular}
\end{minipage}
\captionsetup{font=tiny}
\caption[]{ Sampled points for the singlino-dominated points for Scenario I/II (left/right)  for either $m_{H_1}=125$ GeV or $m_{H_2}=125$ GeV in the reduced SI-SD cross section plane. The color coding corresponds to the singlino content of the lightest neutralino. The vertical and horizontal dashed line show the lower limit on the SI and SD cross section expected for the future experiment DARWIN for a  neutralino mass of about 100 GeV.  Points in the lower left quadrant are below the "neutrino floor",  which are only possible within Scenario II (right-hand side), since they require a  singlino purity  above 99\%. Such pure singlinos are only possible for values of $\lambda/\kappa$ below $\sim$0.03/0.01, as we discussed before \cite{Beskidt:2016egy}. }
\label{f7}
\end{center}
\end{figure}

Most of the sampled points for the chosen scenarios will be within reach of the future direct DM searches. The  comparison of the reduced cross section with the expected future sensitivity of DM experiments on the cross section, for which we take the proposed DARWIN experiment as an example, shows in Fig. \ref{f4} that in parts both, singlino- and Higgsino-dominated LSPs can be out of reach of future experiments. The Higgsino-dominated LSPs can be out of reach mainly because of the high coupling to Higgs bosons, which reduces the relic density, thus leading to a small reduced scattering cross section by the small value of $\zeta \approx 10^{-4}$, as shown in Fig. \ref{f3-3}.  

Singlino-dominated LSPs can be out of reach because of the small coup\-ling to SM particles and thus small  scattering cross section, which may be reduced even further by the  factor $\zeta$ for a relic density being below the observed relic density (dark blue points). Points with a low SI and SD cross section have a large singlino component, as demonstrated in Fig. \ref{f7}.
Here the singlino-dominated points for Scenario I/II (left/right)  for either $m_{H_1}=125$ GeV or $m_{H_2}=125$ GeV are shown in the reduced SI-SD cross section plane. The color coding corresponds to the singlino content of the lightest neutralino. The vertical and horizontal dashed line show the lower limit on the SI and SD cross section from the future experiment DARWIN for a WIMP mass of about 100 GeV, which is close to the "neutrino floor". 
Scenario I will be fully covered by future direct dark matter experiments, while for Scenario II points with a singlino purity of about 99\% will evade detection in the future. Such high purities are only possible for small values of $\lambda/\kappa$ below $\sim$0.03/0.01, in which case the lightest neutralino is decoupled and interacts weakly with SM particles. 
At the same time the annihilation cross section for the correct relic density is still fulfilled by the sum of many co-annihilation channels with the second-lightest and third neutralino $\tilde{\chi}_2^0/\tilde{\chi}_3^0$ and lightest chargino $\tilde{\chi}_1^\pm$ for neutralino masses around 100 GeV. In this case the lightest chargino, the second-lightest and third neutralino are all of the order of $\mu_{eff} \approx 100$ GeV.

\section{Conclusion}
We surveyed the cross sections for the SI and SD dark matter searches in the semi-constrained NMSSM.
The parameter space was sampled by considering a space spanned by  the 7  Higgs masses, which reduces to a 3-D  space, if one takes into account that one Higgs mass has to be equal to the observed Higgs mass of 125 GeV and the heavy Higgs bosons are practically mass-degenerate. The advantage of projecting on the space spanned by masses is that the masses are largely uncorrelated  and one can marginalize over the highly correlated couplings.  From the sampling in the mass space we obtained the range of the neutralino masses and the corresponding SI and SD cross sections, as shown in Fig. \ref{f4} for two different ranges of allowed couplings corresponding to Scenarios I and II.

While Scenario I with large $\lambda/\kappa$ couplings can be explored by the SI and SD searches, the new scanning technique reveals also that significant regions of the NMSSM parameter space in Scenario II cannot be explored with projected experiments, even if one considers both, SI and SD, searches. 
Such scenarios, which cannot be explored, are not evident from previous investigations \cite{Xiang:2016ndq,Cao:2016cnv,Ellwanger:2016sur}. Since the singlino content of the LSP in these scenarios, displayed  in the left bottom quadrant of the right panel of Fig. \ref{f7}, is above 99\%, they cannot be explored  by the LHC either.

\section*{Acknowledgements}
Support from the Heisenberg-Landau program and the Deutsche Forschungsgemeinschaft (DFG, Grant BO 1604/3-1)  is warmly acknowledged.

\bibliographystyle{lucas_unsrt}
\bibliography{paper8}

\end{document}